\documentclass{elsart} 

\usepackage[dvips]{graphics}
\usepackage{epsfig}
\usepackage[figuresleft]{rotating}

\usepackage{amssymb}
\usepackage{amsmath}

\newcommand{\oversim}[2]{\protect{\mbox{\lower0.5ex\vbox{%
   \baselineskip=0pt\lineskip=0.2ex
   \ialign{$\mathsurround=0pt #1\hfil##\hfil$\crcr#2\crcr\sim\crcr}}}}} 
\newcommand{\df}{${\sf df}$}

\newcommand{\tcr}{t_{\rm cr}}
\newcommand{\bgal}{{\sc buildgal}\,} 
\newcommand{\leftb}{<\!\!} \newcommand{\rightb}{\!\!>}
\newcommand{\D}{\displaystyle} 
\newcommand{\toto}{{\sc MaGalie}}

\newcommand{\boldv}[1]{\ifmmode \mbox{\boldmath $ #1$} \else 
 \mbox{\boldmath $#1$} \fi}
\newcommand{\rmd}{\ifmmode \:\mbox{{\rm d}}\else \mbox{ d}\fi }

\def\astrobj#1{#1}
\def\url#1{{\ttfamily\def\/{/\discretionary{}{}{}}#1}}

\begin{document}

\begin{frontmatter}

\title{Efficient N-body Realisations of axisymmetric Galaxies and Haloes}
\author{Christian M. Boily\thanksref{email1}}
\thanks[email1]{Email : cmb@ari.uni-heidelberg.de}
\address{Astronomisches Rechen Institut, M\"onchhofstrasse 12-14,
D-69120 Heidelberg}
\author{Pavel Kroupa\thanksref{email2}}
\address{Institut f\"ur Theoretische Physik und Astrophysik,
Universitaet Kiel, D-24098 Kiel, Germany}
\thanks[email2]{Email : pavel@astrophysik.uni-kiel.de} 
\author{Jorge Pe\~{n}arrubia-Garrido\thanksref{email3}}
\thanks[email3]{Email : jorpega@ari.uni-heidelberg.de}
\address{Astronomisches Rechen Institut, M\"onchhofstrasse 12-14,
D-69120  Heidelberg}


\begin{abstract}
We present an efficient method for building  equilibrium 
multi-component galaxies with non-spherical haloes and bulges. The 
gist of our approach is to  tailor the velocity ellipsoid directly to the
 geometry of the mass distribution. Thus we 
avoid computing the anisotropic velocity dispersions which leads to 
large savings in the cpu budget. The computational time of the 
algorithm for triaxial equilibria scales linearly 
with the number of particles, $N$. The approximate solution to the velocity 
field causes structural relaxation: tests with $N = 50,000$ show 
that fluctuations of the inertia tensor 
 (not exceeding the 10\% level) disappear within one half of a revolution at 
the half-mass radius. At later times equilibrium  properties settle to values 
close to those sought from the initial conditions.  
 
 A disc component is then added as described by Hernquist (1993). 
Incorporating the above algorithm to his code \bgal, test 
runs  show that the stability of the disc 
 against vertical heating is not substantially modified by using 
our method. 

The code, \toto, is made generally available$^1$. 
\end{abstract}


\end{frontmatter}

\section{INTRODUCTION} 
\label{sec:intro}
\noindent The visible \astrobj{Milky Way} and other disk galaxies may be contained in 
extended dark-matter haloes of largely unknown morphology.  
Spherical haloes are typically assumed, but there are 
indications that this matter component may be significantly flattened,
or even triaxial (see Zaritsky 1998, Sackett 1999, Rusin \& Tegmark 2000). 
An important indicator is the non-isotropic
distribution of satellites around disc galaxies, 
or the Holmberg effect (see Holmberg 1969). This may be
a result of dynamical friction of non-spherically symmetric dark
haloes acting on
the satellites (Pe\~{n}arrubia, Boily \& Kroupa  2000).  \\

Equilibria of self-gravitating  galaxies are achieved for a given  
distribution function (\df) based on three integrals of the motion. 
This approach faces a difficulty in that only the  first two classical 
integrals (total energy $E$ and one component of angular momentum $\boldv{J}$) 
are known, the final integral taking an assumed form  (see e.g. Lupton, Gunn 
\& Griffin 1987; Einsel \& Spurzem 1999). A greater difficulty still is that  
galaxies are heterogeneous systems of several components, 
and hence  are hardly represented adequately with a unique \df. A self-consistent equilibrium may be constructed by resorting to 
 numerical integration of the equations of motion. 
Barnes (1988) introduced  an iterative scheme by
which particles relax in a fixed potential representing a galaxy
component. This component is then given a particle representation in 
near-equilibrium. \\

Hernquist (1990, 1993) has introduced a by-now standard scheme 
which solves moments of the collisionless Boltzmann equation 
using tailored distribution functions. Three-component galaxies so
constructed have the advantage over the Barnes approach of  
closer agreement with  the sought  equilibrium. 
In practice his algorithm is highly
efficient  when embedding a galactic disc in spherical
components. (Benchmarking examples will follow in subsequent
sections.) However we have found that for the purpose of constructing 
 axisymmetric (non-spherical) bulges and haloes, the scheme requires 
significantly more cpu-time, mainly because one must solve numerically for 
the components of the velocity dispersion  locally in three
directions. This introduces an $N^2$ scaling of the computational time. 
By contrast, in spherical symmetry the velocity dispersion
is known immediately from a one-dimensional integral (cf. [\ref{eq:vsquare}]
below).  
The algorithm of Hernquist (1993) remains efficient for up to tens of thousand 
of particles, however  the computational time becomes prohibitive for larger numbers. 
 Given that particle numbers adequate to galactic systems 
by far  out-stretch available
computer power, algorithmic problems such as this require solutions. 

In this contribution we by-pass the Boltzmann equations approach, 
 by noting that galactic components follow a well-defined hierarchy of 
sizes and masses. This  suggests
 a perturbative treatment of particle potentials from 
the largest scales, down.  Equilibrium solutions to 
Boltzmann equations in spherical symmetry are first transformed 
to the desired morphology directly, by  mapping the velocity
ellipsoid to the density isocontours. We  construct 
 equilibrium multi-component  axisymmetric 
galaxy models by perturbing the velocity field of individual components 
to account for the added background potential (or, embedding). 
 The computational time required for the whole procedure scales with
 particle number as for  spherically symmetric models. 
 Thus multi-component axisymmetric 
equilibrium configurations with millions of halo particles are
easily constructed. In the following Section we review the problem  
and motivate our approach. This is then implemented and tested using 
a grid code. Benchmarking and comparisons with Hernquist's standard 
\bgal follow. We conclude with 
 suggestions for applications and further improvements.

\section{METHOD}\label{sec:meth}
\subsection{Scaling of Computational Time} 
\noindent  Approximate 
solutions based on moments of the Jeans equations have proved useful when 
constructing equilibrium galaxies. Hernquist (1993) has introduced a scheme 
whereby equilibrium galaxy components are first constructed in isolation, then zipped together by adding the gravity  of the other components in turn. 
The modified binding energy of individual particles is matched by 
a similar increase in kinetic energy to preserve the 
structure of each component.

The case when all components are spherically symmetric allows a simplification.
Then the added binding energy is measured
 from the increase in mass inside a
particle radius $r$. To add together two or more spherical components only requires to sort the particles radially. 
The computational time required by a scheme  such as {\sc quicksort} 
scales no faster than the $3/2$-power of particle number (see
e.g. Press et al. 1992, \S 8). 

For non-spherical potentials however, the modified velocity
field (and square velocity dispersions) 
requires the computation of the local potential (and its spatial derivatives) 
for each particle. The exact, direct-summation algorithm to do this requires a 
number of operations increasing in proportion to $\propto N^2$, the
square particle number. 
The particle realisation of the density itself requires $\propto N$
operations. 
Thus in general we may write the total computational time to 
realise a compound equilibrium of $c$ components as 

\begin{equation}  
T = \sum_{i=1}^c\left(\overbrace{N_i\, t_{den,i}}^{\rm Density} + \underbrace{N_i\, t_{vel,i}\ {\rm or}\, N_i^2\, t_{vel,i}}_{\rm Spherical\, or\, aspherical} + \overbrace{N_i\, \sum_{j=1}^c  N_j\, t_{cor,i}}^{\rm Background\, potentials} 
\right)\end{equation} 
 where $N_i$ is the number of particles for component $i$; 
 $t_{den}, t_{vel}$ the computational times per particle to realise 
 the density and velocity fields for respective 
components; and $t_{cor,i}$ the characteristic time  for adjusting the velocity field 
of $i^{th}$-component  particles due to non-spherical potentials. 
 For three-component galaxies made of a spherical bulge (b), disc (d) and aspherical halo (h),  we would write 

\[ T  =  N_h\, t_{den,h} + N_h^2\, t_{vel,h} + N_h\, (N_d + N_b)\, t_{cor,h}
+ N_d\, t_{den,d} + N_d^2\, t_{vel,d}\, + \]
\begin{equation} \ \ \ \  N_d\, (N_h + N_b) \, t_{cor,d}
+ N_b\, t_{den,b} + N_b\, t_{vel,b} +  N_b\, (N_h + N_d)\, t_{cor,b} \ .
\label{eq:scaling} \end{equation} 
Clearly massive haloes call for large particle numbers and so the leading 
terms in (\ref{eq:scaling}) are $\propto N_h^2$ and $\propto  N_h\, N_d$. 
 We wish to remove these dependencies on $N_h$. 

In any multi-component galaxy, the largest-scale 
structure would be closest to equilibrium self-gravity. 
This suggests that we look for a general scheme whereby the relative masses and
scales of components are taken into account. Figure~\ref{fig:schema}
 provides a cartoon 
representation of our approach. Since details of the small scale components 
are lost on the much larger ones, a correction to the velocity field 
of larger structures would only involve truncated expansions of the 
small-component potentials. By contrast, a light structure requires more 
accurate treatment of the surrounding mass distribution: this is shown  
as light and heavy  arrows on Fig.~\ref{fig:schema}. We adopted this strategy 
 by solving for the self-gravitating aspherical halo through a  
map of  the velocity field of the 
spherical solution.  As we shift down the scale ladder to smaller structures, higher-order expansions provide accurate corrections to the velocity 
field of individual components. Thus feedbacks from disc and bulge on the halo 
velocity field are given a low-oder series-expansion treatment,
while the velocity field of disc and bulge particles are 
 adjusted with a high-order expansion of the 
 halo potential. Below we take a stepwise angle to implementing this 
algorithm. 


\subsection{Self-gravitating axisymmetric equilibria} 
\noindent  The basic step  consists 
in building axisymmetric systems from a transformation of 
spherically symmetric equilibria. If we consider as given the geometric aspect 
ratio and the mass profile of the equilibrium sought, it is then sufficient 
to construct  a velocity field to match the gravity.  
In cylindrical coordinates $(R=\sqrt{x^2+y^2},z, \theta)$, the 
orbits of stars in an axisymmetric harmonic potential are recovered from integrating
two decoupled equations of motion. In the case of a uniform oblate spheroid 
of density $\rho$, axes $a,c\ ; a > c$, eccentricity $ e = \sqrt{1 - c^2/a^2}$, the 
potential inside the bounding volume is (Table~2-1 in Binney \& Tremaine 1987, hereafter BT+87; Chandrasekhar 1969, \S 3) 

\begin{equation} \Phi_{sp}(R,z) = - \pi G\rho \left( {\sc I}(e) a^2 -
A_1 (e) R^2 - A_3 (e)z^2 \right) \label{eq:potential_inside} \end{equation}
with known algebraic expressions for ${\sc I}, A_i$. The motion of a
test particle is found from 
solving for 

\begin{equation} \ddot{x}_i = - \nabla_i \Phi = - 2\pi G\rho A_i x_i\ , i = 1 ... 3 \label{eq:harmonic} \end{equation}
 which admits a periodic solution in each component. 
 Averaging the square velocity and position over one cycle, 
and substituting for $A_i$ from Table 2.2 of BT+87, we obtain

\begin{equation} \frac{\leftb v^2_3 \rightb}{\leftb v^2_1\rightb} = 
\frac{A_3}{A_1} \frac{\leftb x^2_3\rightb}{\leftb x^2_1\rightb} =
 \frac{A_3}{A_1} \left( 1 - e^2 \right) =  
2 \frac{a/c - \sin^{-1}(e)/e}{\sin^{-1}(e)/e - c/a}\ 
\frac{c^2}{a^2}\ .
\label{eq:toto1} \end{equation} 
From (\ref{eq:potential_inside}) we find lines of constant potential intersecting the $z=0$ plane and the z-axis such that for these coordinates 
$ A_1 x_1^2 = A_3 x_3^2$.
 Defining the eccentricity of isopotential contours 

\begin{equation} e_{\Phi} \equiv \sqrt{\left(1-\frac{x_3^2}{x_1^2}\right)_\Phi}
 = \sqrt{1 - \frac{A_1}{A_3}}\ \label{eq:e_phi} \end{equation}
we may relate $e_\Phi$ to the geometric eccentricity $(e)$ and that of 
the velocity ellipsoid, $e_v$, using (\ref{eq:toto1}) and
(\ref{eq:e_phi}) 

\begin{equation} 
e_v \equiv \sqrt{ 1 - \frac{\leftb v^2_3\rightb}{\leftb v^2_1\rightb} } = \left[ \frac{e^2-e_\Phi^2}{1 - e_\Phi^2} \right]^{\half}\ . \label{eq:e_v}
\end{equation}
The three quantities satisfy $ e_{\Phi} < e_v < e $, and hence the velocity 
ellipsoid is never as flat as the mass distribution that gives rise to it. 
 This result may be derived  directly from the virial theorem applied to the 
system as a whole (see BT+87, \S 4.3). Therefore, for the harmonic potential, (\ref{eq:e_v}) applies
equally to averaged quantities and individual orbits. 

In practice galaxies and haloes show peaked density profiles and hence 
the exact result (\ref{eq:e_v}) has limited application. An alternative 
to (\ref{eq:e_v}) is obtained by considering a star on a loop orbit in the mean potential 
of all the others. Since the matter is concentrated around the centre of 
gravity, a series expansion will be adequate when the orbit avoids the central region. Consider therefore the 
potential exterior to a homogeneous spheroid, of axes $(a,c)$ equal to the mass-weighted means of the peaked distribution.   
This is written in a form analogous to (\ref{eq:potential_inside}) (BT+87, Chandrasekhar 1969),  

\begin{equation} \Phi_{sp}(R,z) = - \pi G\rho \frac{a^2c}{(a')^2c'} \left( {\sc I}(a',c') (a')^2 - \sum_i A_i (a',c') x_i^2 \right)\ . \label{eq:potential} \end{equation}
 In terms of $(a,c)$ we have 

\begin{equation} \frac{R^2}{(a')^2}+ \frac{z^2}{(c')^2} = 1\ ;\ a' \equiv \sqrt{ a^2 + \lambda }, \ c' \equiv \sqrt{ c^2 + \lambda } \ .\label{eq:primed_axes} \end{equation} 
For oblate spheroids the solution 
 $\lambda(a,c,r) $ is expressed in algebraic form:  

\[ 2\ \lambda = r^2- a^2 (2-e^2) + \sqrt{ (a^2-c^2)^2 + 2 a^2 e^2 ( z^2 - R^2 ) + r^4 } \ , \]
where $r^2 = R^2 + z^2$. (A relation similar to this one also exists 
for prolate spheroids but is omitted for reasons of clarity.) 
The solution $\lambda > 0$ applies outside the bounding surface of the spheroid. For a loop orbit with $r > a $ a series expansion 
of (\ref{eq:potential}) leads to (cf. Goldstein 1980, eq. 5-88) 
 
\begin{equation} \Phi_{sp} (R,z) = -\frac{GM_o}{r} + \frac{GM_o}{2r^3} ( I_z - I_R )
 P_2(\cos\theta) + {\sc O}([a/r]^4) \label{eq:quadrupole} \end{equation}  
where $\tan\theta = z/R$, $P_n(x)$ is a Legendre polynomial, 
and $I_x$ an eigen-component of the inertia
tensor per unit mass and we consider only the half-space $z>0$  in
what follows. Integrating the equations of motion for the potential (\ref{eq:quadrupole}) leads 
to solutions of the form 

\begin{equation} x_i (t) = x_{o,i} \cos[ w_1 (t-t_o) ]\ \left( 1 + \frac{w^2_2}{w^2_1} \frac{x^2_{o,i}}{8 r^2} \cos^2 [ w_1 (t-t_o) ] + ... \right) \label{eq:anharmonic} \end{equation} 
where 

\[ w_1^2 = G\overline{\rho}  + \frac{9}{2} G\overline{\rho}\, \frac{I_z - I_R}{r^2}\ ; \ w_2^2 =  \frac{9}{2} G\overline{\rho}\, \frac{I_z-I_R}{r^2}\ ;\ \overline{\rho} \equiv \frac{M}{4\pi r^3/3} \ . \]
Thus when $r \gg a > c $,  $w_2\rightarrow 0$ with $1/r^2$ since the $I_i$'s
 are constants. The orbits  describe
ellipses of the same eccentricity in both \boldv{\dot{x}} and \boldv{x}. Note 
when $I_z = I_R$, (\ref{eq:anharmonic}) reduces to solutions for the
right-hand side in  (\ref{eq:harmonic}). 
For such orbits we would therefore set 

\begin{equation}  e_v = e \ . \label{eq:e_v.eq.e} \end{equation} 
Our solutions (\ref{eq:e_v}) and (\ref{eq:e_v.eq.e}) bracket the range of 
possible aspect ratios the velocity ellipsoid may assume in equilibrium. 
Realistic galaxy equilibria will fall somewhere between strict homogeneity and
point-mass distributions. We therefore sought a sensible interpolation 
 for $e_v$ in the range $e_\Phi$ to $e$ to account for this. 
The number of possibilities  is {\em a priori} endless. 
We chose for each particle $i$

\begin{equation} e_{v,i}^2 = e_\Phi^2 + ( e^2 - e_\Phi^2 ) \sqrt{ 1 - \frac{\leftb r^2\rightb}{r_{g,i}^2} } \label{eq:interpol} \end{equation} 
which was found to yield adequate equilibria. 
In computing $e_v$, we substitute in the above 
the mass-weighted average $r^2$ inside the spherical volume $4\pi r_{g,i}^3/3$. To obtain an 
eccentricity for the velocity ellipsoid that reflects the character of the 
gravitational field 
 sampled by the star over one orbit, we compute the gravitational radius 

\[ r_{g,i} = \frac{GM}{-E_i} \ , \] 
where $E_i$ is the binding energy of the $i^{th}$ star, and $M$ 
the total mass of the system. 
 The velocity ellipsoid computed from (\ref{eq:interpol}) is the same for all stars of 
the same binding energy. We give some motivation for the interpolation 
(\ref{eq:interpol}) through an example. 

To fix ideas, consider a  
Dehnen model for spherical galaxies. Dehnen (1993) introduced a family of 
potential-density pairs which provide a suitable fit to bulges and 
cuspy ellipticals : 
 
 \[ M(r) = M_o \left( \frac{r}{r+r_c} \right)^{3-\gamma} \, \, , \]
 \begin{equation}
 \Phi(r) = \frac{GM_o}{r_c}\, \times  \left\{ \begin{array}{lr} 
\frac{\D 1}{\D 2-\gamma} \left( 1 - \left
[ \frac{\D r}{\D r+r_c}\right]^{ 2-\gamma}\right)\, \,&  (\gamma \neq 2)  \\ 
 \ln \frac{\D r}{\D r+r_c} \, \, & (\gamma = 2) \end{array}\, \right. \, . 
\label{eq:dehnen} \end{equation}
Here $(r_c,\gamma)$ are two parameters and the total system 
mass  $= M_o$. The case $\gamma = 1 $ corresponds to a spherical Hernquist
(1990)  density profile.  Let $r_c = 1/10$, and we truncate the distribution  
at $r = 1 $ for convenience. With $M_o = 1.21 .. $ the mean square radius 
  $\leftb r^2\rightb = 0.349 ... $, enclosing  63\% 
of the mass. We invoke the virial theorem and  set $ E_i = -\Phi_i/2 $
for a given stellar orbit. Using this in  (\ref{eq:dehnen}) we compute 
$r_{g,i} = G M_o / (-\Phi_i/2) = 2 (r_i+r_c)$; the 
ratio of averaged values $ \leftb r_g\rightb/\leftb r\rightb = 1/4$. 
If we now squash the sphere down 
the z-axis to achieve an aspect ratio $c/a = 1/2$, we compute
from (\ref{eq:interpol})  an eccentricity $ e_v \approx 0.860 ..\
$, close to $ e = \sqrt{1-c^2/a^2} = 0.866 .. $.  Setting 
$\gamma = 0 $ so that the central region is of uniform density, we
find from  (\ref{eq:e_v}) 
for this model $ e_v = 0.832 .. $, nearly unchanged from the previous 
value. This is because 
 only $\approx 12\%$ of the mass now resides inside $r = r_c$. 
Thus unless the galaxy shows a 
broad uniform density region, the solution (\ref{eq:interpol}) on the mean 
  lies  close to $e$.

\section{Stand-alone Axisymmetric Haloes and Bulges} 
\noindent We now implement the scheme described above. We begin by 
summarising the steps used in constructing spherical equilibria. 
Applications will centre on the Dehnen models (\ref{eq:dehnen}), as they 
cover a range of central density gradients (through the index $\gamma$) 
 from which to test the method.  We concentrate on two aspects, 
 namely the stability of equilibria so constructed and the range of 
cases to which this  algorithm may 
be applied. We consider isolated self-gravitating haloes first;
multi-component models are discussed in the next section. 

\subsection{Spherically Symmetric Equilibria} 
\noindent  For the  case of spherically symmetric mass profiles the 
potential at radius $r$ is 

\begin{equation} \Phi (r) = - \int_{\infty}^{r}\frac{GM(x)}{x^2} \rmd x = 
  - \int_{\infty}^{r} G \frac{\rmd x}{x^2} \int_0^{x} 4\pi u^2\rho(u)\rmd u
\label{eq:phi} \end{equation}  
provided the run of mass $M(r)$ increases more slowly than $\propto r$ at large
distances. 
The mass distribution is a given quantity of the problem. A particle 
representation of any mass profile $M(r)$ can be obtained
straightforwardly, first by filling a spherical volume uniformly 
by Monte Carlo method, then 
mapping  the mass shells to the desired profile, i.e. $M_h(\hat{r}) =
M(r)$, with $M_h$ the uniform-density spherical profile. 
The solution $r(M)$ 
 is not known in analytic form in general and must be tuned by
interpolation when only a discrete representation $M_i(r_i)$ is given
on input. One advantage
over solving directly by Monte Carlo method for $r(M)$ is that 
a unique  realisation  $M_h(\hat{r})$ may be transformed to a variety of
profiles. This essentially keeps any bias due to $\sqrt{N}$ 
fluctuations the same 
for different $M(r)$. This is  especially important 
 in systems where the effect of low-N statistics may be felt, such 
as in central galactic cusps. Interpolation errors would wipe out this 
advantage when the desired mass profile is known only on a coarse mesh 
$\{r_i\}$.

The problem is complete once the velocity 
field satisfying the   Jeans equations under $\Phi(r)$ is solved for. 
 From the  first of these equations (cf. Binney \& Tremaine 1987, \S4.2), 

\begin{equation}  \frac{\rmd\rho\overline{v^2_r}}{\rmd r} + \frac{\rho}{r} \left[ 2
\overline{v^2_r} - \overline{v^2_\theta} - \overline{v^2_\phi}\right] = - \rho
\frac{\rmd \Phi}{\rmd r}, \label{eq:jeans1} \end{equation}  
for an isotropic velocity field the solution takes the form 

\begin{equation} \overline{v^2_r} = \frac{- 1}{\rho(r)} \int_r^\infty \rho(x)
\frac{\rmd\Phi}{\rmd x} \rmd x \ = \frac{1}{\rho(r)} \int_r^\infty
 \rho(x) \frac{G M(x)}{x^2} \rmd x\ . \label{eq:vsquare} \end{equation}  
 With a known  velocity dispersion at each radius it is customary to
 seek closure by picking a form for the velocity \df. 
We will assume a  velocity  dispersion taking locally a
 Maxwellian form (Hernquist 1993), 

\begin{equation} F_v (\boldv{v},r) {\rmd d}^3\boldv{v} = 
4\pi \left(\frac{1}{2\pi\sigma^2}\right)^{\frac{3}{2}} \ v^2 \exp( - v^2/2\overline{v^2_r}
 ){\rm d}v \ . \label{eq:maxwellian} \end{equation}  
N-body realisations of a
Maxwellian distribution would require integrating  to infinity in
velocity space. In practice bound particles all have velocities 
below the local escape velocity  $(= \sqrt{-2\Phi(r)})$, effectively
setting an upper-limit for $v$ in (\ref{eq:maxwellian}). A
proper implementation of this renormalises the velocities 
to preserve the Maxwellian character of the field, so 

\begin{equation} \overline{v^2}(r) = 3 \ \sigma^2(r) \ ,\end{equation} 
 where $\sigma$ is the one-dimensional velocity dispersion at
$r$ and $\sigma^2 = \overline{v^2_r}$ for non-rotating distributions.

\subsection{From spheres to axisymmetry} \noindent 
To 
 construct axisymmetric halo and bulge components, we proceeded as follows. 
Our 
starting point is a  spherical equilibrium constructed as in the above. 
With  $M(r)$
known, the potential $\Phi(r)$ and velocity dispersion are evaluated for  
member stars from (\ref{eq:phi}) and (\ref{eq:vsquare}). Each particle
is then given a velocity drawn from  the distribution 
(\ref{eq:maxwellian}). Since only  one-dimensional 
 integrals are involved, the computational requirements  for this 
part of the calculation scales with $\propto N$, where  $N$ is the
total number of particles. 

We then perform a homologous transformation of the position of each particle 
to achieve the desired morphology. The case where isodensity contours are 
concentric oblate spheroids is a good starting point. The particles are 
then mapped according to $(x,y) \rightarrow (x,y)$, $z \rightarrow z \sqrt{1-e^2}$. 
 The quantities $\leftb r^2 \rightb$ and $r_g$ are computed directly 
from $M(r)$, prior to the mapping. 

To modify the velocity field requires evaluating the potential energy of 
each particle in the new spheroidal potential, $\Phi_{sp} (R,z)$. The 
quantity $\sqrt{\leftb r^2\rightb}$ is computed from mass lying inside 
 any particle radius, $r$, 
for which we define a uniform density spheroid of axes 
 $ (\sqrt{\leftb r^2\rightb},  \sqrt{\leftb r^2\rightb}\sqrt{1-e^2})$, total 
mass $M(r)$.  Equations (\ref{eq:potential}) and 
(\ref{eq:primed_axes}) then yield the 
correct potential at $(R,z)$ up to quadrupolar order. By virtue of the virial theorem 
half of the potential energy accrued by the stars through the 
coordinate transformation should be invested in kinetic energy. Thus 

\begin{equation} T_i = E_i - \Phi (r_i)  \rightarrow T_i^\prime = T_i + 
\frac{ \Phi (r_i) - \Phi_{sp} (R_i,z_i)}{2}  \label{eq:deltaek} \end{equation}
for each particle $i$. The velocity vector $\boldv{v}_i^\prime$ must satisfy (\ref{eq:interpol}), which is the case when 

\[ \boldv{v}^\prime_i = \sqrt{ \frac{2 T_i^\prime}{2T_i - v_z^2 e_v^2} } \times ( v_x, v_y, v_z \sqrt{1-e_v^2} ) \ . \]
In the above the primed quantities refer to the transformed, final
 state. The multiplicative factor ensures the new velocity gives 
 the kinetic energy (\ref{eq:deltaek}).           
 The new potential could have been evaluated from (\ref{eq:quadrupole}), by computing the inertia tensor inside the spheroid $(r, r \sqrt{1-e^2})$, 
for each particle. However (\ref{eq:primed_axes}) and (\ref{eq:potential}) 
require only evaluating trigonometric functions and the algorithm is therefore
optimised for the transformation we have performed here. When the 
eccentricity of the isodensity contours vary with position, one  
requires averaged eccentricities in (\ref{eq:interpol}) for every mass shell, 
which is provided through (\ref{eq:quadrupole}) by the tensor of inertia. 

\newcommand{\superbox}{{\sc superbox}\,} 

\subsection{Numerical Setup} 
\noindent The algorithm we have described was tested using the 
 nested-grid integrator  \superbox (Fellhauer et al. 2000).  
 The tests we conducted used in total $N = 50000$ particles. 
Since spherical equilibria provide the work horse of the scheme, we
first constructed a spherical equilibrium to tune up the code. 

The spherical profile used  is the Dehnen 
family (\ref{eq:dehnen}) with $\gamma = 1 $ and $a = 1/10$. The models 
were all truncated at $ R_{cut} = 1$.   
We then evolved the distributions for 3 to 4 circular orbits at the 
half-mass radius, each revolution corresponding to time intervals 
$4\, \tcr = 4 \sqrt{3\pi/16 (G\overline{\rho})^{-1}}$, where 
 $\tcr$ is the dynamical time and 
$\overline{\rho}$ the averaged density at the half-mass radius. Setting $G = 2$
we have $ 4 \tcr = 0.727 .. $ for a single revolution
 about the half-mass volume. At the edge of the system an orbit takes 
 $ \simeq 2.05 .. $ units of time for a revolution.

\superbox  is a
time-centred leap-frog integrator. The timestep we selected $= 0.006
\times\ \tcr $; the radius where
this timestep matches the local dynamical time  for an Dehnen $\gamma = 1$
model is at (roughly) $1\times 10^{-3}\times a$, enclosing $\sim 10^{-4}$
percent of the mass. The dynamics is therefore well resolved in time 
for all particles in the simulations.  
In general the central cusp is poorly resolved by grid codes due to 
finite linear resolution. 
However the three levels of resolution allowed by \superbox  means that 
we can focus on the central region and boost resolution as
needed. Setting the total grid size $= 1/2$ we then find for an 32-mesh grid a linear resolution 
 $l \approx 1/2 \times 1/32 = 0.016$ such that 2  percent  of the total 
mass falls inside the innermost grid points. This would minimise 
 structural evolution, however the coarse central density due to
finite resolution  means that 
 velocity field and potential no longer match there, resulting in a small 
 degree of relaxation at the centre. 

\begin{figure}[t] 
\setlength{\unitlength}{1in} 
\begin{picture}(4.,5.)(0,0) 
\put(-0.25,-1.5){\epsfig{file=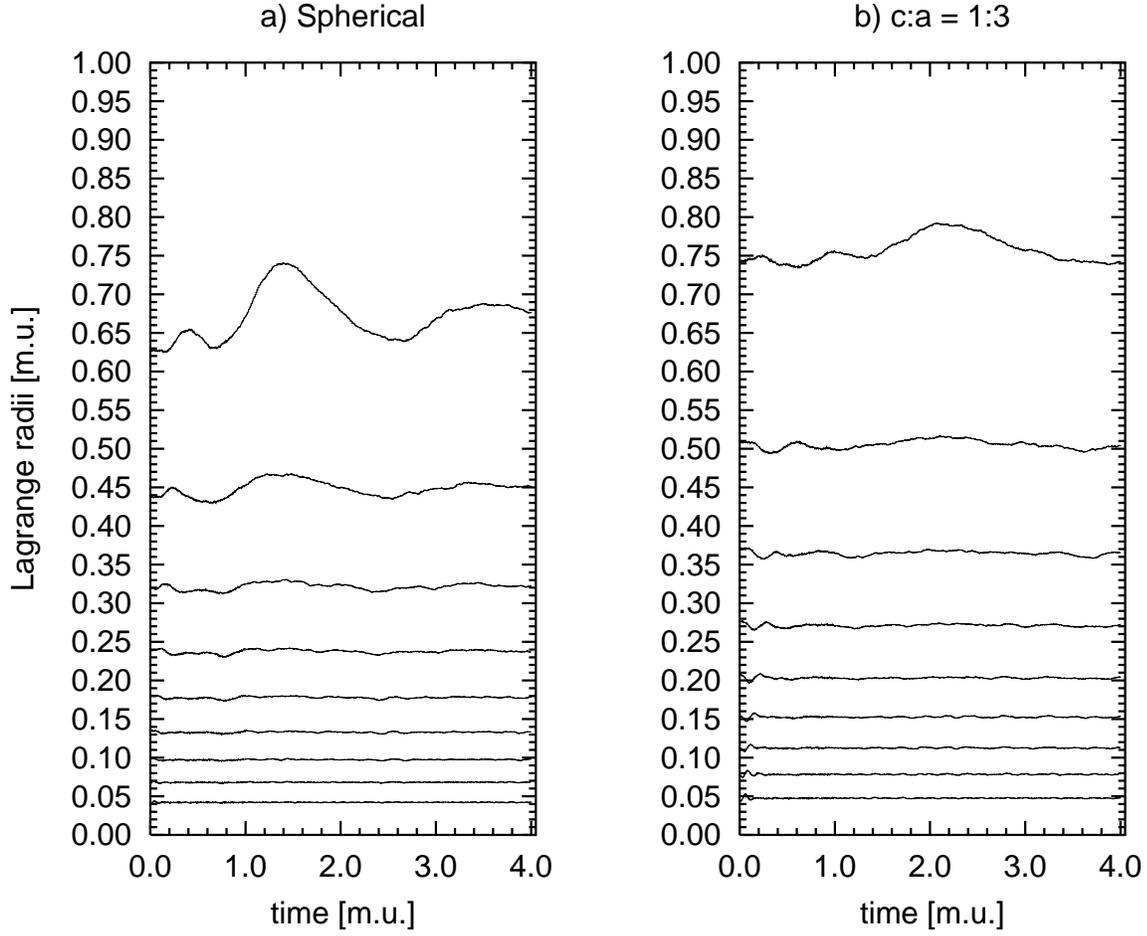,height= .8\textheight,width=.8\textwidth, angle=-0}
             }
       \put(0.,.25){
\begin{minipage}[t]{\textwidth} \caption{\label{fig:dehnen1} 
    (a) Evolution of a spherical Dehnen $\gamma=1$
       model. (b) Same model, but flattened to achieve an aspect ratio $= 1:3$.
 The 10-percentile Lagrange radii are displayed versus time, where 
one unit of time corresponds to a full revolution at half-mass.}  
\end{minipage} 
       } 
\end{picture}  
\end{figure}

The left-hand panels on Fig.\ref{fig:dehnen1} show the  results of evolving a spherical 
Dehnen $\gamma = 1$ model. The figure displays the 10-percentile Lagrange 
radii of the mass sorted in spherical shells. We note that systematic fluctuations are low for the innermost mass shells. However  near the edge appreciable evolution can be seen, at a level of 15\%  for the 90\% mass radius. Bearing in mind this caveat, we then applied the 
same numerical setup to flattened Dehnen models. To preserve the good
resolution of the dynamics in time, we enforced identical central density for
 all models by an homologous transformation of the particle positions,  
  so the dynamical time $\tcr$ remains the same. 

\subsection{Results for axisymmetric models with $\gamma = 1$} 
\noindent Non-rotating 
self-gravitating equilibria of aspect ratios $c/a > 1/3$ are 
 susceptible to  bending modes of instability (Combes et al. 1990,
Sellwood \& Wilkinson 1993). No elliptical galaxy has a projected
aspect ratio flatter than E7 (Binney \& Merrifield 1998). Furthermore
the narrow stream of debris from the Sagittarius dwarf suggests that 
 the halo of the \astrobj{Milky Way} may be as round as or rounder than $E1$
(Ibata \& Lewis 1998). 
Therefore we consider a flattened model 
with $ c:a = 1:3$ as a limiting case  to test the algorithm. This is done
first, before investigating the effects of varying aspect ratio 
and power index $\gamma$. 
 
Figure\ref{fig:dehnen1}(b) graphs the Lagrange radii of an 1:3 flattened Dehnen model. The mass profile has been binned in spherical shells to facilitate 
comparisons with the spherical case. Assessing the overall setup, we find as 
expected a degree of relaxation in the innermost mass bins. Close inspection 
shows fluctuations  of amplitude $\sim 0.01$ forming  at all Lagrange radii. 
These fluctuations propagate inside out, and result from stars leaving the central region. For the model at hand we estimate that $2.8\%$ of the total mass 
 gives rise to them. At the end of this simulation only 0.52\% of stars had 
left the volume of the simulation, therefore we conclude from this exercise 
that $\approx 2\%$ of kinetic energy is not assigned correctly by  
our application of  (\ref{eq:deltaek}) and leads to large errors in
the  binding energy of relatively few particles. 

\begin{figure}[t] 
\setlength{\unitlength}{1in} 
\begin{picture}(4.5,5.)(0,0) 
\put(-.75,-2.25){\epsfig{file=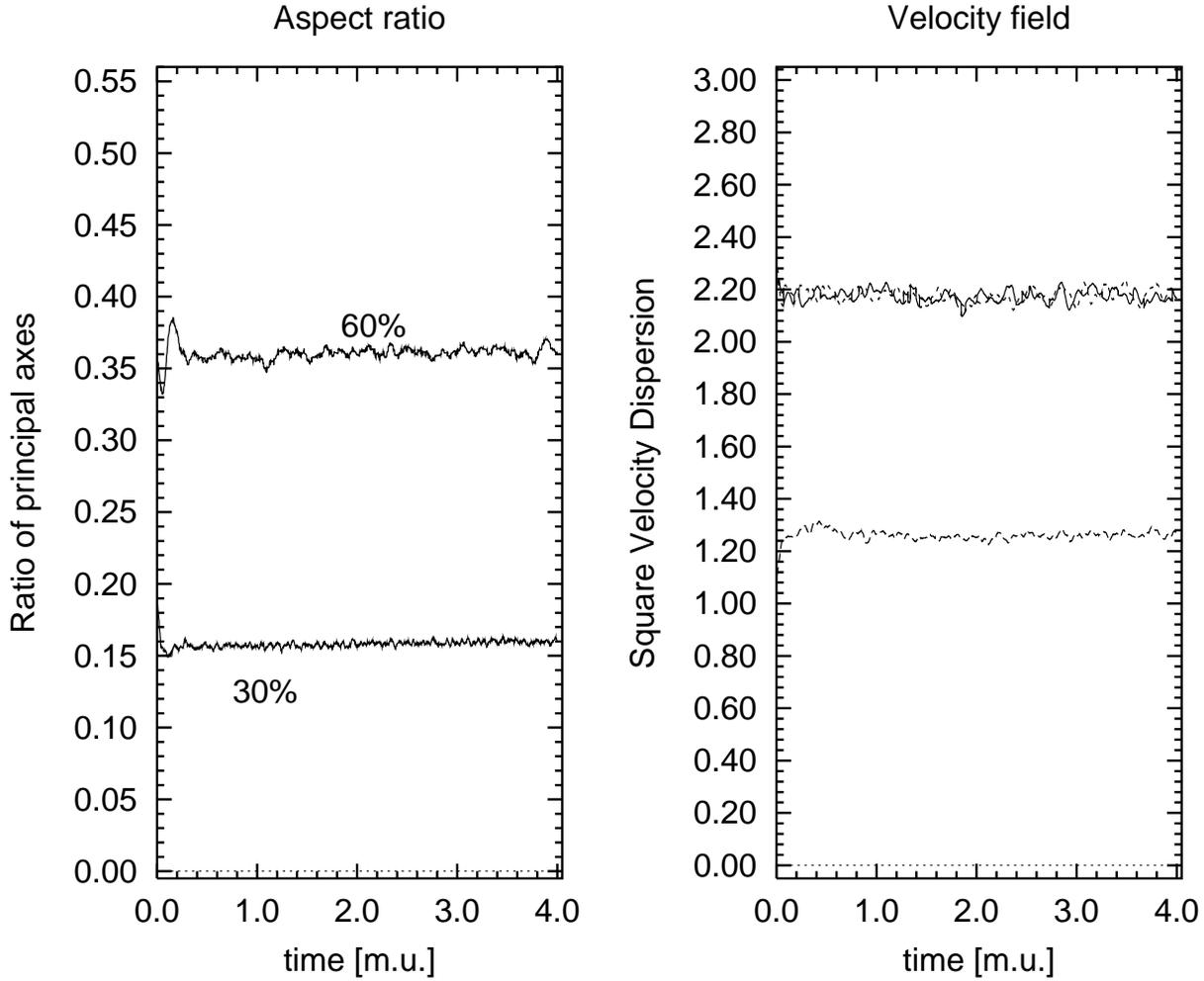,height= 0.85\textheight,width=.85\textwidth, angle=-0}
              }
\begin{minipage}[t]{\textwidth} 
\caption{\label{fig:axes} Left panel: 
        structural evolution of a flattened 1:3 Dehnen $\gamma=1$
       model. The ratio of minor to major axes are computed at
	two different Lagrange radii enclosing 30\% or 60\% of the 
	total mass. Note: the results at 30\% Lagrange radius were 
	divided by two to avoid overlap with the others. Right panel: components of the 
velocity dispersion tensor (cf. Eq. 21) for the innermost 60\%
particles. The lower curve is the z-component. The velocity dispersions were computed along the eigenvectors of the rotational inertia tensor.}  
\end{minipage} 

\end{picture}  
\end{figure}

We monitor in time  the morphology of the model defining the mass-weighted 
principal axes 

\begin{equation} a_i^2 \equiv \frac{ I_j + I_k - I_i }{2} \label{eq:axes}
\end{equation}
where $\{I_i\}$ are the eigenvalues of the inertia tensor and $ \{i,j,k\} = \{x,y,z\}$ 
and permutations thereof (Goldstein 1980, \S 5-3; see also Kroupa
1997). Particles are first  selected in spherical
Lagrangian shells, from which the $I_i$'s are computed. 
A first estimate of the principal axes of the mass distribution 
 is computed from (\ref{eq:axes}). 
 A second selection of particles is then made, with  the
bounding volume now assuming the ellipsoidal shape defined by the
$a_i$'s. The inertia tensor and eigenvalues are recomputed from this 
new selection to obtain the final estimate of the system axes. 
Figure~\ref{fig:axes} (left-hand panel) 
graphs the evolution of the 
quantity $ 2 a_3 / ( a_1+a_2) $ for two mass bins, the 30\% and 60\% Lagrange 
radii. For clarity the 30\% results were divided by a factor two. We find 
the relaxed configuration giving nearly the exact same quantities after 
a half-unit of evolution. It is noteworthy that relaxation was more severe 
in the central region, where the computed aspect ratio drops from 0.18 initially to 0.16 at later times. 
The right-hand panel shows the velocity field for the 
60\% innermost particles. Here there are virtually no signs of evolution or trends 
beyond the level of root-$N$ fluctuations. 

\subsection{Models with $\gamma \ne 1$} 
\noindent We investigated the range of applicability of our approach by 
evolving Dehnen models with different power indices $\gamma$. 
 The solution (\ref{eq:e_v}) for the velocity ellipsoid 
applies to harmonic potentials of uniform density.  Since 
we opted instead to implement (\ref{eq:interpol}), we ask whether equilibria 
can be setup this way with a harmonic central core, i.e. for  models with  
$\gamma = 0$. 

\begin{figure}[t] 
\setlength{\unitlength}{1in} 
\begin{picture}(4.,4.)(0,0) 
\put(-0.5,-1.5){\epsfig{file=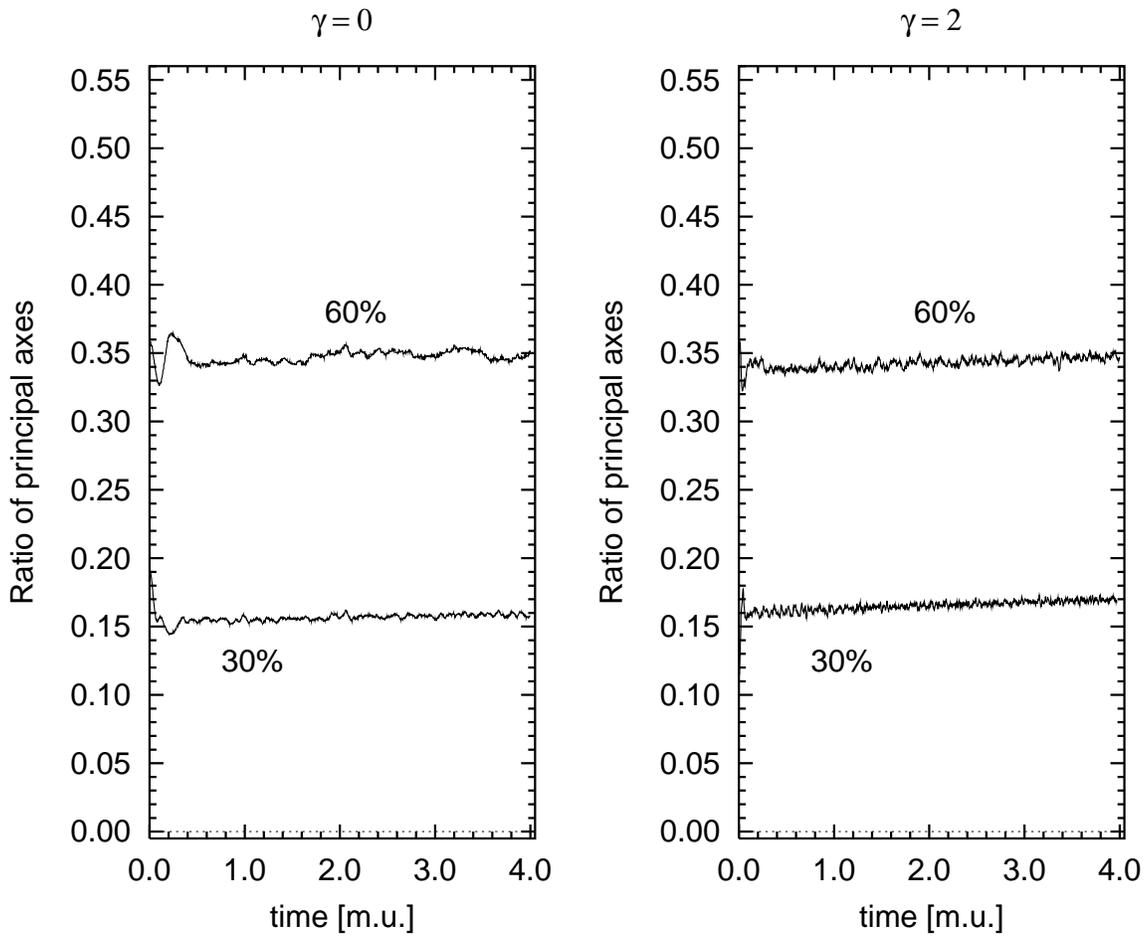,height= .8\textheight,width=.8\textwidth, angle=-0}
             }
       \put(0.,.5){
\begin{minipage}[t]{\textwidth} \caption{\label{fig:flat_axes} 
        As for Fig.\ref{fig:axes}. 
Time evolution of two flattened Dehnen models of aspect ratio 1:3, 
	with $\gamma=0$ (left) and $\gamma = 2$ (right). The principal axes 
(\ref{eq:axes}) were computed at the 30\% and 60\% spherical Lagrange 
radii. The 30\%-results were divided by two to distinguish the two curves
on each panels. 
       }  \end{minipage} 
       } 
\end{picture}  
\end{figure}

Figure~\ref{fig:flat_axes} (left-hand panel) illustrates the impact of the central harmonic core. We have graphed the same quantities as on Fig.~\ref{fig:axes}, using the same numerical setup to validate the comparison. The oscillations 
of the 60\% axis ratio seen early on in the simulation compare in magnitude to 
those observed for the $\gamma = 1$ case. However in the inner region the 
 ratio varies from 0.18 initially, to 0.15 after one time unit of evolution, or down 17\%, somewhat  more than what was observed for the same 
volume in the case of $\gamma = 1$, where variations  were of  $\simeq 11\%$ relative magnitude.

The multipolar expansion of the gravitational field suffers less from truncation if the matter is more concentrated about the centre of gravity. Therefore 
 Jaffe's (1983) profiles  (Dehnen $\gamma = 2$) provide another test since 
they are more concentrated. The right-hand panels on Fig.~\ref{fig:flat_axes} show indeed that 
fluctuations due to dynamical relaxation are much reduced for the period $t < 1/2$.
 The principal axes ratio fluctuates between $0.17\pm0.01$ or 5.9\% for the inner 30\% mass radius. At larger volume it fluctuates initially by a similar relative
fraction. We would expect this property to prove helpful when
 modelling dark haloes as isothermal gas, 
for which asymptotically $\rho \propto r^{-2}$ at large distances. 


\section{Multi-component Models} 
\noindent The map (\ref{eq:interpol}) provides equilibria for 
one-component self-gravitating spheroidal galaxies. We now generalise to 
multi-component galaxies. Our goal is to verify that the embedding of new 
components does not introduce further  evolution in the structure of the 
galaxy. In an effort to bring out algorithmic, as opposed to dynamical, 
biases, we have put together a bulge-disc-halo model galaxy where the bulge 
and halo have the same  spherical Dehnen $\gamma = 1 $ mass profiles. We  then 
opted to combine  a spherical 
bulge, exponential disc and axisymmetric halo, requiring respectively zero-, high- and low-order corrections to the monopole for a precise determination  
of  the potentials. We modified  a version of \bgal kindly provided to us 
by L. Hernquist and  based on  his 1993 algorithm. We implemented 
the map of Section 2.2 for axisymmetric galactic potentials, 
taking into account the hierarchy of individual components. 
 
\subsection{Disc, bulge \& halo galaxy} \noindent 
 While not attempting to model the Galaxy, we set up model galaxies with 
 relative masses and lengths in rough agreement with those of 
Freudenreich (1998) for the \astrobj{Milky Way} 
 (see Sackett 1997, Binney \& Merrifield 1999, \S 10). In model units, the 
bulge has a mass and radius $= 1$; the core length $r_c = 0.10$. 
The exponential disc has a radial scale length $h = 1.0$, 
scale height $z_o = 0.3$, total radius = 10 and mass = 3.0. 
We chose scale lengths $h$ and  $z_o$ 
 large compared with the linear resolution $l = 0.016$ in the
inner region of our  grid code. 

 The heavier spheroidal 
halo was  given an 1:2 aspect ratio as described  in Section 2.2. We took 
$z$ as the symmetry axis so the spheroid equator lies in the x-y plane.  
The halo has a total radius = 20.00, mass = 30.00, and we chose a halo core 
length $r_c = 2.0$. The halo and disc masses are in the ratio 4:1 at the 
edge of the disc, so the halo drives the dynamics there 
but not overwhelmingly so. 
This is done on purpose, with a view to perturb the halo equilibrium, as it
must adjust to the disc gravity. 

\subsection{Computing the potentials} \noindent 
Experience  and pragmatic considerations led to the following simplifying 
tricks. When computing the disc feedback on embedded bulge particles, we have 
kept the direct-summation algorithm from the original code, since the dimensions of the bulge are comparable to the disc scale length. Hence a multipole 
expansion would  require high-order series expansion which we deemed 
not essential, as the number of bulge particles is less or comparable
 to the 
number of disc particles, which already gives $t \propto N_d^2$
scaling of computational time (cf. equation \ref{eq:scaling}). We
found the direct-summation computation, contributing a term $\propto
N_d N_b$ in the total computational time, gives better result than a
expanding the disc potential to high-order, although a crude
evaluation using (\ref{eq:potential}) also provides a sensible
equilibrium - see below.  The feed-back of the bulge on disc particles however is  accounted 
for already when  the monopole term of its potential is added to that of 
the disc. 

The halo and disc interactions require some care. We found it sufficient 
for an axisymmetric halo to expand its potential to quadrupolar term
(cf. [\ref{eq:quadrupole}]) 
and adding this to the disc's (and the bulge's) self-gravitating potentials. 
Equation (\ref{eq:quadrupole}) is easily differentiated in three dimensions.
 The kinetic energy is then added to the disc and bulge particles for 
each degree of freedom in the same proportion as the components of the 
halo gravitational forces (to account for the work done in reshaping
the system). In this way, relatively more kinetic energy is 
invested in the x-y plane, when compared with the z-direction, to fight off 
the enhanced gravity due to the flattened halo in the equatorial plane of the
spheroid. 
As the rms disc aspect ratio is close to 1:12, the boost in disc 
kinetic energy required to maintain equilibrium  was added to the circular 
motion at cylindrical radius $R$ for each disc particle. Since the 
streaming motion at $R$ is known, energy can be directly added to each 
particle once (\ref{eq:deltaek}) has been computed. 

As halo particles spend little time near or in the plane of the disc, we 
treated the disc as a highly flattened oblate spheroid, for which the 
potential is computed from (\ref{eq:potential}). To do so 
 we averaged over the vertical structure
 of the disc, however the original scales of height and radial 
decay are preserved if we use as spheroid axes the mass-weighted disc 
parameters $\leftb R\rightb$ and $\leftb z \rightb$. This effectively removes
the term $\propto N_d\, N_h$ of the total computational time (\ref{eq:scaling}).

\subsection{Results} 
\begin{figure}[t] 
\setlength{\unitlength}{1in} 
\begin{picture}(4.,5.)(0,0) 
\put(-0.35,.5){\epsfig{file=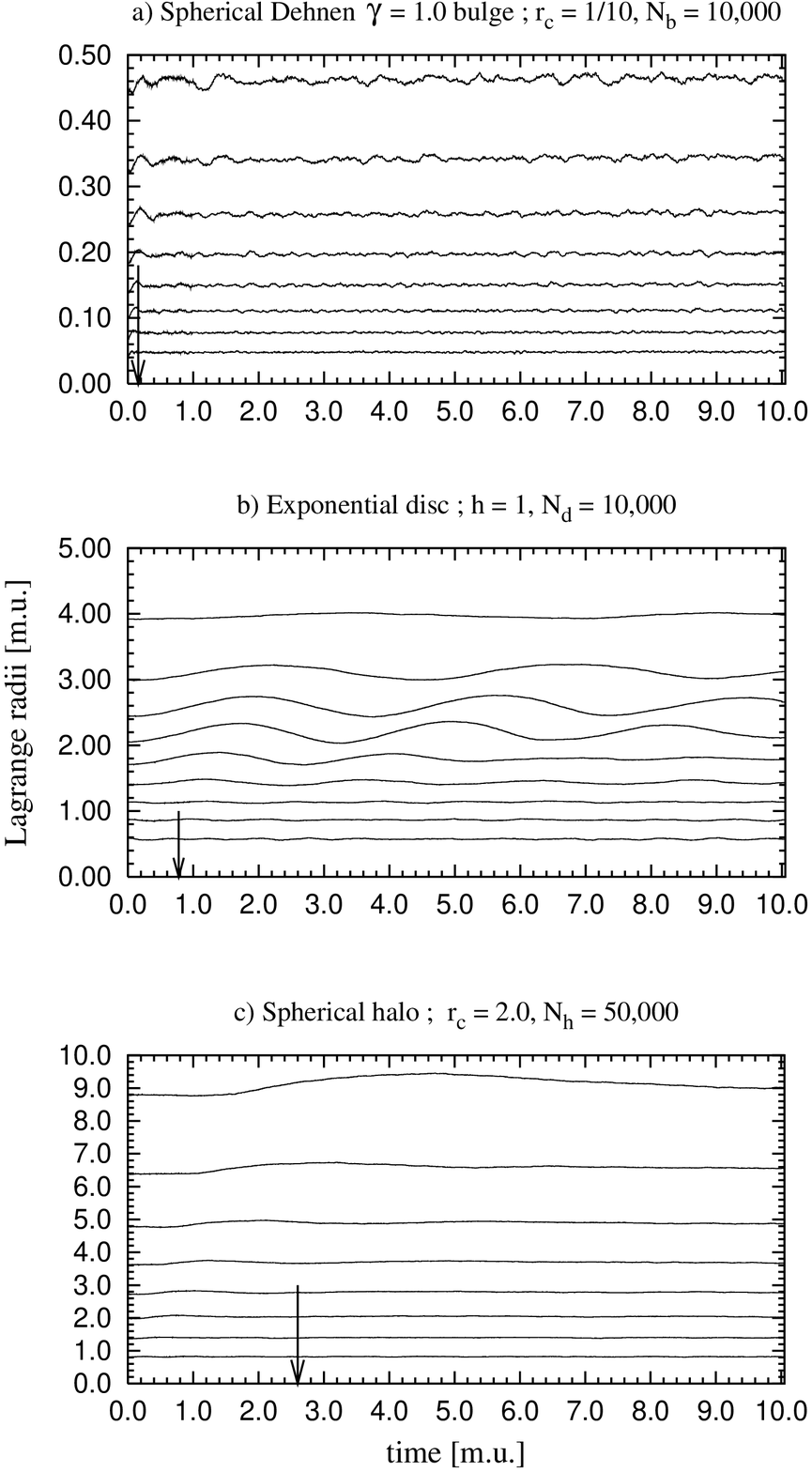,height= 0.55\textheight,width=0.55\textwidth, angle=-0}
             }
\put(3.2,.5){\epsfig{file=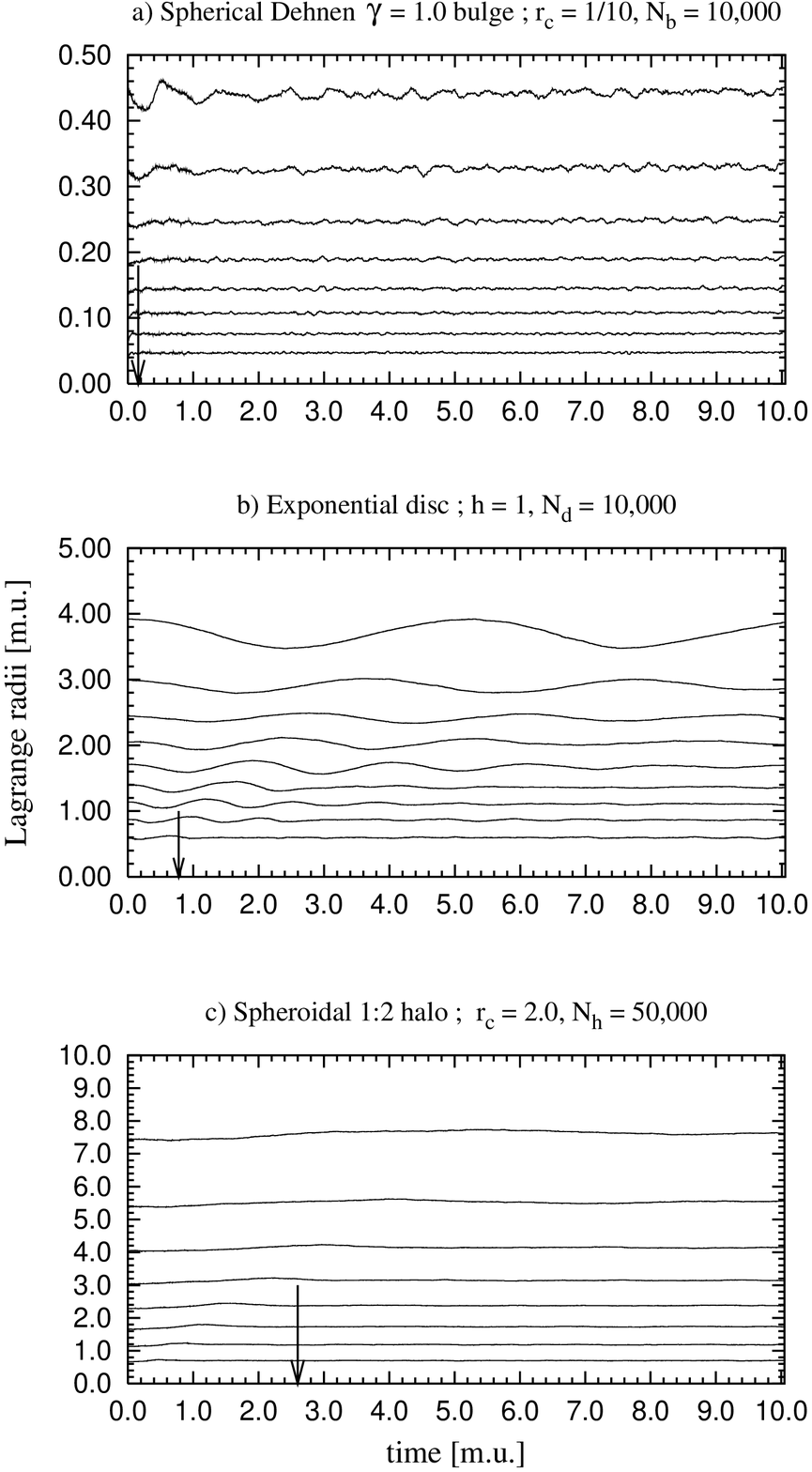,height= 0.55\textheight,width=0.55\textwidth, angle=-0}
             }
       \put(0.,.25){
\begin{minipage}[t]{\textwidth} \caption{\label{fig:3com..sphere} Left panels: 
  evolution of the Lagrange radii (including 10, 20, 30 ... percent of 
 the component's mass) 
 for a three-component model with spherical bulge and halo. 
Right panels: as on the left, but with a spheroidal halo of aspect ratio 1:2. The arrows indicate the circular period at the 50\% Lagrange radius (bulge and halo) and at the radial scale length $h$ (disc, middle panel).  }  
\end{minipage}  
                   } 
\end{picture}  
\end{figure}

\noindent 
For comparison purposes we display two models on Fig.~\ref{fig:3com..sphere}.
 The first one 
has both spherical bulge and halo, while the second has an oblate spheroidal 
halo of aspect ratio 1:2. The motivation for doing this is clear: for spherically symmetric components, the multipole approach becomes exact, since only 
the monopole term contributes to the potential of each component. Models of a
disc galaxy with spherical bulge and halo thus 
  minimise relaxation effects. 

The structure of the bulge is well 
accounted for by sorting the mass on spherical Lagrangian shells. 
For the case of the disc, of mean aspect ratio $\simeq 1:12$, the Lagrangian 
shells effectively follow the cylindrical radii. 
Contrasting the two models, with and without spherically symmetric halo, 
we find little effect on the radial structure of disc and bulge introduced by our approximate treatment. In both the cases, radial oscillations are seen 
to propagate from the inside, outwards, while taking several dynamical times 
to dampen. On the figure the arrows indicate the dynamical time $\tcr$ at the half-mass 
radius (bulge and halo) or at $R = h$ (disc) of individual components 
in isolation. These radial oscillations give
rise to transient patterns on visual inspection. We note that after a
few 
 revolutions of the disc at $R=h$, much of the fluctuations have disappeared or are much attenuated. 

\begin{figure}[t] 
\setlength{\unitlength}{1in} 
\begin{picture}(4.,5.)(0,0) 
\put(-0.35,.35){\epsfig{file=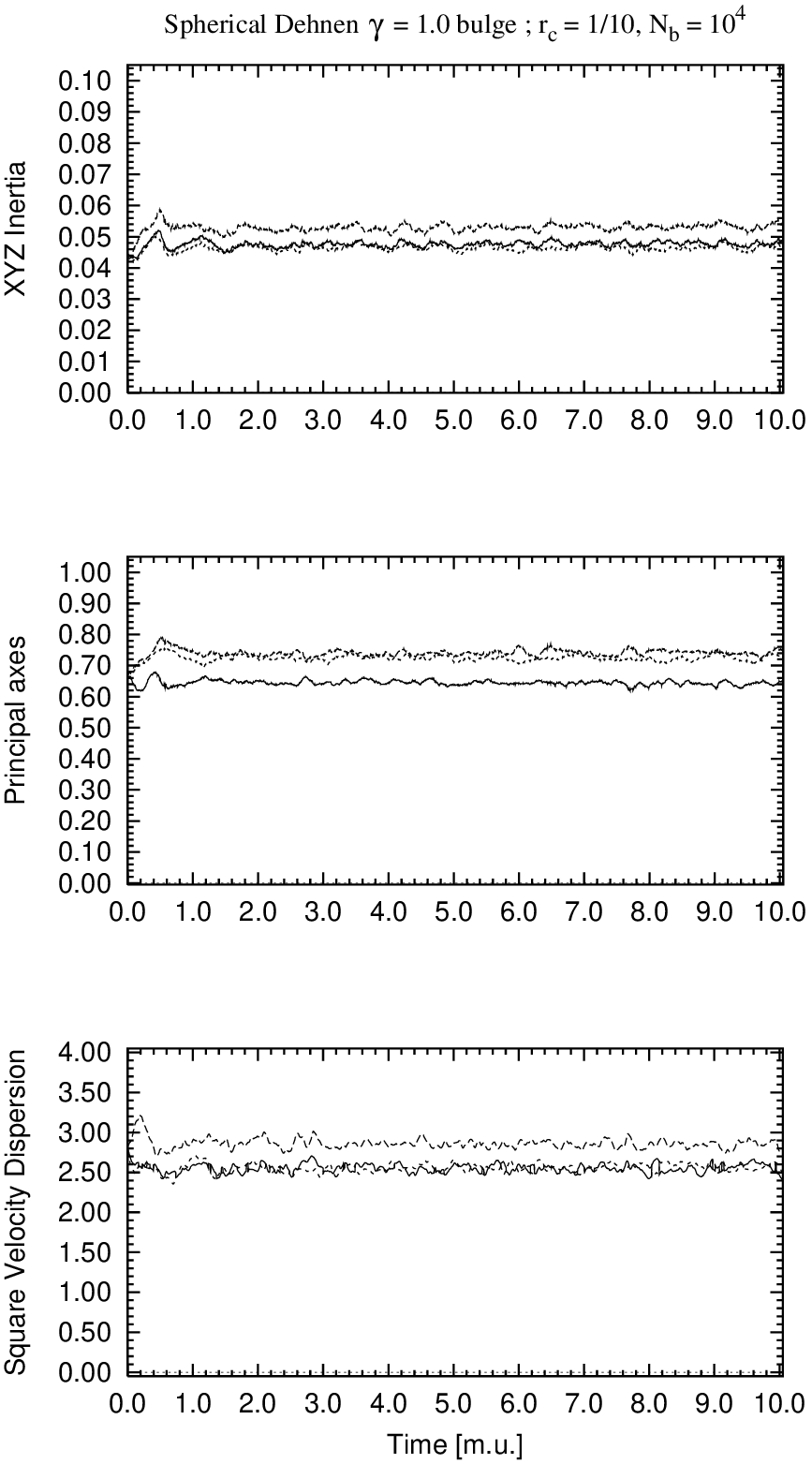,height= 0.555\textheight,width=0.5\textwidth, angle=-0}
             }
\put(3.2,.35){\epsfig{file=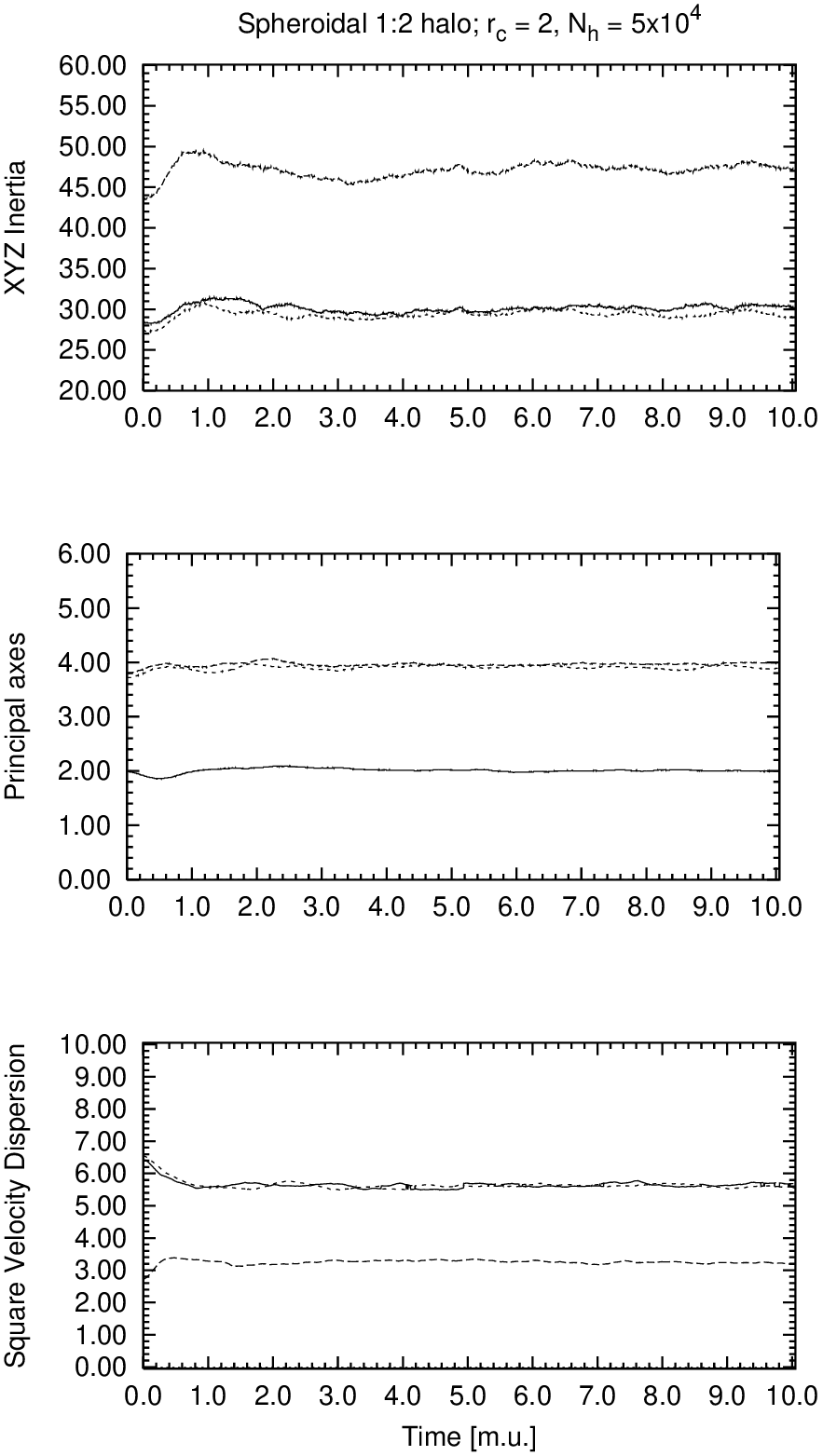,height= 0.55\textheight,width=0.5\textwidth, angle=-0}
             }
       \put(0.,.2){
\begin{minipage}[t]{\textwidth} \caption{\label{fig:3com..tensors} 
  Time-evolution of the morphology of individual components. Left-hand panels: 
 central bulge parameters; right-hand panels: halo parameters. The quantities were measured 
at the 50\% spherical Lagrange radius. The principal axes were computed as in 
(Fig.~\ref{fig:axes}). The XYZ Inertia are the eigenvalues of the rotational inertia tensor
(Goldstein 1980).} 
\end{minipage}  
                   } 
\end{picture}  
\end{figure}

The moments of inertia give  insight into the structure of individual galaxy
 components. 
 On Fig.~\ref{fig:3com..tensors} we display the moments of inertia, principal 
axes and velocity dispersion components of the bulge and halo for the case 
where the halo is an oblate spheroid. We find as on Fig.~\ref{fig:dehnen1} 
 slight changes in the ratio of principal axes of the halo. However, after 
 settling, the structure remains close to the one specified on input. The 
bulge, on the other hand, is clearly stretched out of spherical symmetry, 
due to the combined pull of the disc and halo: at the edge of the bulge, $R = 1$, the halo mass $\approx 10/3$ is already several times that of the bulge, 
which is not fully self-gravitating anymore. The aspect ratio of the bulge in 
equilibrium is close to $0.65/0.75 = 0.87$, when we would want unity. The 
external forces acting on the bulge leads to expansion, as seen from the 
increased moments of inertia (top left-hand side panel). A quick solution 
to counter this effect would be to stretch the bulge initially to
 mildly prolate structure of axis ratio $ a:c \sim 1:1.1$, so as to
 relax to near sphericity. We have not resorted to this therapy 
and chose instead to illustrate the limitations of the algorithm. 

Turning to the vertical structure of the disc, we averaged the height $|z|$ 
of stars inside $R = 2h$, since this region is well resolved by our grid 
code and errors due to numerical integration are reduced. We computed 
both average vertical height and the one-dimensional velocity
dispersion $\sigma_z$ for ten snapshots covering ten time units of
evolution, or  
four revolutions. Table~\ref{tab:zdisc} gives the results. Early in the 
evolution of the disc, the vertical structure shrunk by $\approx 15\%$, as 
the mean height drops from $\approx 0.2$ to $\approx 0.17$. We observed 
a decrease of the standard deviation as well, of the same magnitude, which 
leads us to conclude in a global contraction of the vertical disc structure. 
This is matched by a rise in velocity dispersion of the same relative
magnitude (see Table~\ref{tab:zdisc}). Since $<v_z>\approx0$ and 
is much less than the vertical velocity dispersion, 
the bulk motion remains negligible thoughout.

\subsection{Benchmarks} 
\label{bench}
\noindent Approximate methods such as the one we have presented would not be
 of interest without a handsome pay-out at the computer. We find  the scaling 
properties of the method with a series of numerical tests. 
Specifically, we first bench-marked the original version of the code \bgal for 
spherical bulge + disc galaxy models. We then constructed an axisymmetric 
bulge using the full algorithm, which computes the response of the disc 
to bulge potential exactly. Since bulge and halo are constructed in the same
way in \bgal, and the disc embedding done in the same exact fashion, 
we may relabel the bulge as halo. This is so when  referring to \bgal below. 
In a second set of tests, we repeated these experiments, using  the 
new version of the code.   
 The same default disc parameters were used in all our tests, and haloes were 
given Hernquist (1990) mass profiles with $r_c = 0.1$, and truncated at the disc edge. 

 All benchmarks refer to 
 an 333 MHz G3 Apple processor operating under  S.u.S.E. Linux 6.4 OS; we 
 compiled \bgal with the widely available 
GNU Fortran compiler {\sc g77 -O4}. The performance were done with 
 disc and bulge particle numbers, $N_d, N_b$, 
in the range $5,000$ to $5\times 10^4 $ (disc); 
and $5000$ to $N_b = 5\times 10^5$ (bulge or halo). 

We solved for the operation times $t_{x,i}$ in (\ref{eq:scaling}), first 
as a set of linear equations by constructing stand-alone disc and bulge (no 
feedback terms present), then re-doing the calculations with disc and bulge 
together. Our results are listed  in Tables~\ref{tab:scaling} and
\ref{tab:scaling2}. 
The most important point to notice is the large increase in computer time for models with 
an axisymmetric bulge, which scales effectively with the square of particle 
number. The turn-over to quadratic dependence is approximately $2\,
10^4$ particles. The validity of (\ref{eq:scaling}) is inferred from the  convergence 
of the operation times $t_{x,i}$'s with increasing $N$. We have extrapolated to 
infinite particle numbers using the polynomial fit routine {\sc
pzextr} of Press et al. (1992). The numbers are truncated to
the last significant digit according to 
error estimates obtained from the results of the two highest-order 
polynomial fits. 

Repeating this exercise with the new algorithm, we concentrated on the case of
an axisymmetric halo + disc, since the case of a spherical halo would give 
essentially the  same $t_{x,i}$'s. We therefore  have to compute only the 
extra computational time required for the transformation of the halo to
 axisymmetry and its coupling
 with the disc. We find as expected that the time required to adjust the 
disc and halo velocity fields scale in proportion to their respective 
particle numbers. A comparison of total computational time shows that 
axisymmetric compound models require only a modest increase in computer 
time when compared with spherically symmetric components. 
The new code, \toto\ (for Mache Galaxie), can thus construct 
galaxies with bulge, disc and halo, all in axisymmetry, at virtually 
no extra costs.

\section{CONCLUSIONS}
\label{sec:concl}
\begin{figure}[t] 
\setlength{\unitlength}{1in} 
\begin{picture}(4.,5.)(0,0) 
\put(.40,1.){\epsfig{file=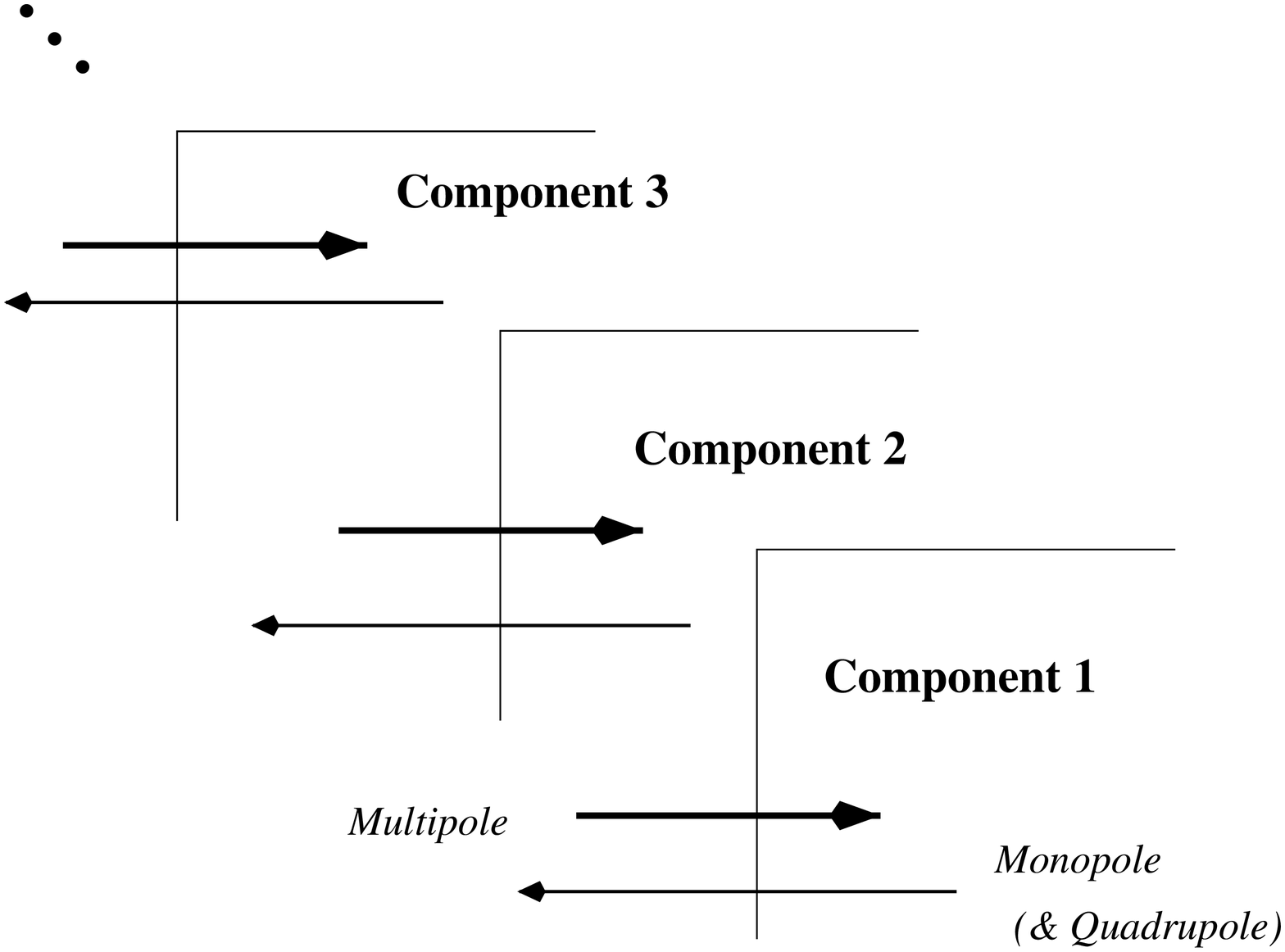,height= .6\textheight,width=.6\textwidth, angle=-0}
             }
       \put(3.2,6.0){
\begin{minipage}[t]{.5\textwidth} \caption{\label{fig:schema} 
    Coupling galaxy components of different linear sizes and masses. A low 
accuracy treatment of the total potential 
is sufficient to maintain the largest, most massive component in equilibrium, 
while it induces a relatively stronger response from the smaller 
components, requiring 
higher-order treatment of the potential for these components. The 
strength of the feedback is shown here as light and heavy arrows.}
\end{minipage} 
       } 
\end{picture}  
\end{figure}

\noindent We have presented a method of constructing approximate equilibria 
in good agreement with those sought. In particular the algorithm scales 
linearly with bulge and halo particle numbers, when these are axisymmetric
or spherical. The approach takes into account implicitly 
 the relative scales of components involved, and treats the interactions between components by expanding the potential 
to required order (Fig.~\ref{fig:schema}). 

The errors introduced from treating non-spherically symmetric components was
found to be no worse than the case where only spherical components surround 
a disc. The required computational time for this algorithm 
scales  with particle numbers as 

\begin{equation} t_{\rm cpu} = N_h\, t_{den,h} + N_h\,t_{vel,h}  +
N_h\, t_{cor,h} + N_d\, t_{den,d} + N_d^2\, t_{vel,d} + N_d\,
t_{cor,d} \label{eq:last} \end{equation} 
where the parameters $t_{x,i}$ are given in Table~\ref{tab:scaling2}
 for a G3 processor  (see Section~\ref{bench}). 
With this algorithm non-spherical haloes surrounding galaxies can be
constructed as a matter of routine. The results of a parameter survey
of orbital decay in aspherical haloes will be presented in a separate 
contribution (Pe\~narrubia et al. 2000). 

We have limited our discussion to prolate spheroidal 
potentials. The approach is readily applicable to prolate or  
triaxial structures however. To construct triaxial equilibria,
 one would map the axes of the velocity ellipsoid 
as in (\ref{eq:interpol}) to the geometry of the triaxial body. Equations 
 (\ref{eq:potential}) and (\ref{eq:quadrupole}) are easily 
generalised to triaxial systems (BT+87). Likewise, net rotation of individual 
components is implemented by alignment of particle angular momenta (see e.g. Lynden-Bell 1962, Hernquist 1993). Models of barred galaxies may possibly 
be constructed 
by converting random into bulk motion of a triaxial component, but we have 
not implemented this. 

 The disc
self-gravity now represents the bottleneck through the $N_d^2$ term in equation (\ref{eq:last}).   For situations where
 the disc  substructure can be neglected, this dependence may
be removed by  constructing a  disc from flattening a 
spheroid to small aspect ratio, then converting random motion to 
 provide rotational support. When the disc structure matters to the 
dynamics however, this approach  is 
unlikely to yield sensible disc equilibria, because the vertical and 
radial  motions in thin structures require fine-tuning the velocities 
locally, hardly in line with such a  brute-force approach. Note however that 
one may construct a disc similar to the \astrobj{Milky Way}'s by considering the 
thin and thick discs as separate components.  These may then be
combined in the fashion described in Section 2. 
Sellwood \& Wilkinson (1993) discuss methods of constructing stable thin galactic 
discs comprehensively. 


\vskip 10mm
\noindent{\bf Acknowledgements} \vskip 3mm
\noindent CMB and JPG acknowledge support from the German government through
an SFB 439 grant at the Karls-Ruprecht-Universit\"at Heidelberg.  

\noindent{\bf Note added in proof -} After this article was submitted,
we found a natural, quick-fix solution to the `$N_d^2$'
scaling of the thin disc construction algorithm (cf. Tables 2 and 3): 
it consists  
in adding together $m$ realisations of an $n$-particle disc to reach 
the desired total particle number $N_d = m \times n$. The total cpu 
then scales with $m \times n^2 \sim n\times N_d$, or linearly with $N_d$. 
 Though $n$ and hence the coefficient of proportionality should be 
reasonably  large to dampen numerical noise, 
the approach is well-suited to parallel algorithms. 


\newpage 
\begin{table}[h]
\begin{center} 
\begin{tabular}{rcccrc}
 time & $N$ & $ <|z|>$ & $\delta z$ & $<v_z>$ & $\sigma_z^2$ 
 \\\hline  
      &           &        &       &  & \\ 
  0   &      5921 & 0.207  & 0.176 & $-8.14\times10^{-3}$ & 0.912 \\ 
  1   &     5833  & 0.169  & 0.163 & $4.6\times10^{-4}$  &  1.084\\
  2   &  6147 &  0.169     & 0.160 & $1.74\times10^{-2}$  & 1.032\\
  3   &    5960 & 0.168 &  0.162 &  $4.04\times10^{-2}$ & 1.061\\
  4   &    5936 &  0.167 & 0.158 &  $4.79\times10^{-2}$ & 1.072\\
  5   &   6006  & 0.168 &  0.164  & $1.86\times10^{-2}$ & 1.070\\
  6   &    5895&  0.169    &  0.161 &   $-4.5\times10^{-4}$ &  1.068\\
  7   &     5975&  0.171 &    0.165 &   $4.09\times10^{-2}$ & 1.060\\
 8     &   5895&  0.170 &  0.158  &  $8.44\times10^{-2}$ & 1.078\\
 9     &   5930 & 0.173   &  0.163  &  $8.47\times10^{-3}$ & 1.066 \\
 10    &    5973 & 0.170 &   0.159 &   $4.29\times10^{-2}$ & 1.099 \\
\end{tabular}
\end{center} 
\caption{Time-evolution of the vertical structure of an embedded disc in 
a flattened spheroid. Only particles inside cylindrical radius $R = 2 h$, where
$h$ is the radial disc scale length, are included when computing mean height 
and velocity dispersion. The standard deviation $\delta z \equiv \sqrt{<\!\!(|z|-<\!\!|z|\!\!>)^2\!\!>}$. }
\label{tab:zdisc}
\end{table}

\begin{table}[h]
\begin{center} 
\begin{tabular}{rcccrccrr}
\multicolumn{9}{l}{(A) Total cpu = $N_h \times ( t_{den,h} + t_{vel,h} ) +
2\,N_d\,N_h\, t_{cor,h} + N_d\, t_{den,d} + N_d^2\, t_{vel,d} $ } \\  \hline \\ 
$N_h$ & $t_{den,h}$ & $t_{vel,h}$  & $t_{cor,h}$ & $N_d$ & $t_{den,d}$ & $t_{vel,d}$ & \multicolumn{2}{c}{CPU [seconds] } \\
& $[10^{-5}$ s] & $[10^{-5}$ s] & $[10^{-5}$ s] & & $[10^{-5}$ s]& $[10^{-5}$ s] & True & Computed\\
$5\,10^3  $ & $  3.80  $ & $6.70$ & $ 0.096 $ &$5\, 10^3 $ & $27.2 $ &$0.368$ &141.9 & 83.7\\ 
$  \,10^4  $ & $3.90$ & $6.10$ & $0.092$ & $5 \,10^3 $ & $27.2$ & $0.368$ &186.4   & 141.1\\
$  \,10^4  $ & $ |  $ & $ |  $ & $0.096$ & $  \,10^4 $ & $27.3$ & $0.222$ &417.7   & 231.4\\
$  \,10^4  $ & $ |  $ & $ |  $ & $0.106$ & $2 \,10^4 $ & $27.4$ & $0.159$ &1066.5   & 862.0\\
$  \,10^4  $ & $3.90$ & $6.10$ & $0.107$ & $5 \,10^4 $ & $27.3$ & $0.112$ &3884.6  & 3653.9\\
$2 \,10^4  $ & $3.80$ & $4.90$ & $ - $ & $  \,10^5 $ & $26.8$ & $0.102$ &10228.5 & 10027.9\\ 
$5 \,10^4  $ & $3.80$ & $4.10$ & $     $ & $ $ & &  &3.95 & 3.64\\
$  \,10^5  $ & $3.80$ & $3.80$ & $     $ & $ $ & &  &7.60 & 7.28\\
$5 \,10^5  $ & $3.70$ & $3.60$ & $     $ & $$  & &  &36.50&36.40 \\
${\infty}$ & $\boldv{3.64}$ & $\boldv{3.54}$ & $\boldv{0.114}$ &$\infty$ & $\boldv{26.5}$ & $\boldv{0.100}$ & \\
\\ 
\multicolumn{9}{l}{(B) Total cpu = $N_h\, t_{den,h} + N_h^2\,t_{vel,h}
+ 2\,N_d\,N_h\, t_{cor,h} + N_d\, t_{den,d} + N_d^2\, t_{vel,d} $ } \\ 
\hline \\ 
$N_h$ &  $t_{den,h}$ & $t_{vel,h}$ & $t_{cor,h}$ & $N_d$ & $t_{den,d}$ & $t_{vel,d}$ & \multicolumn{2}{c}{CPU [seconds]} \\
& $[10^{-5}$ s] & $[10^{-5}$ s] & $[10^{-5}$ s] & & $[10^{-5}$ s]& $[ 10^{-5}$ s] & True & Computed  \\ 
$5 \,10^3$ & $ 12.4 $ & $(0.120)$& $0.096$ & $5 \,10^3$ & $27.2$ & $0.368$ &172.0 &110.4 \\  
$  \,10^4$ & $ 12.4 $ & $0.115$  & $0.092$ & $5 \,10^3$ & $27.2$ & $0.368 $ &301.6 &247.6 \\
$  \,10^4$ & $ |    $ & $ |   $  & $0.096$ & $ \,10^4 $ & $27.3$ & $0.222$ &532.9 & 437.9\\
$  \,10^4$ & $ |    $ & $ |   $  & $0.106$ & $2 \,10^4$ & $27.4$ & $0.159$ &1181.7 &968.5 \\
$  \,10^4$ & $ 12.4 $ & $0.115$  & $0.107$ & $5 \,10^4$ & $27.3$ & $0.112$ &3999.9 & 3760.5\\ 
$2 \,10^4$ & $ 12.5 $ & $0.108 $ & $-  $ & $  \,10^5$ & $26.8$ & $0.102$ &10661.3 &10453.0 \\
$2 \,10^4$ & $ 12.5 $ & $0.108 $ & $ $ & & & &434.5 & 426.5\\    
$5 \,10^4$ & $ 12.4 $ & $0.107 $ & $ $ & & & &2681.2 &2656.9 \\ 
${\infty}$ & $\boldv{12.4}$ & $\boldv{0.106}$ & $\boldv{0.114}$ &$\infty$ & $\boldv{26.5}$ & $\boldv{0.100}$ &  &  \\
\end{tabular}
\end{center} 
\caption{Benchmarks of the code \bgal for (a) spherical halo + disc
and (b) axisymmetric halo + 
disc models. Notation refers to (\ref{eq:scaling}); $t_{cor,d} =
t_{cor,h}$ by symmetry and has been omitted in the table. We computed
CPU values from (\ref{eq:scaling}) using the asymptotic limits shown
in bold face; true CPU corresponds to clocked time.}\label{tab:scaling} \end{table}

\begin{table}[h]
\begin{center} 
\begin{tabular}{rcccrcccrr}
\multicolumn{10}{l}{Total cpu = $N_h\, t_{den,h} + N_h\,t_{vel,h}  +
N_h\, t_{cor,h} + N_d\, t_{den,d} + N_d^2\, t_{vel,d} + N_d\,
t_{cor,d}$ } \\ \hline \\ 
$N_h$ &  $t_{den,h}$ & $t_{vel,h}$ & $t_{cor,h}$ & $N_d$ & $t_{den,d}$ & $t_{vel,d}$ &  $t_{cor,d}$ & \multicolumn{2}{c}{CPU [seconds]} \\
&$[10^{-5}$ s]  & $[10^{-5}$ s] & $[10^{-5}$ s] & & $[10^{-5}$ s]& $[10^{-5}$ s] & $[10^{-5}$ s] 
& True & Computed \\ \\
$5 \,10^3 $ & $3.60$ & $9.40$ & $1.10$ & $5\, 10^3 $ & $13.4$ & $ 0.137$ & $1.80$ &35.7 &20.2 \\
$  \,10^4 $ & $3.80$ & $6.90$ & $1.20$ & $5\, 10^3 $ & $|$    & $ | $    & $|$ &36.2 &20.7 \\
$5 \,10^4 $ & $3.80$ & $4.80$ & $1.30$ & $5\, 10^3 $ & $|$    & $ | $    & $|$ &40.0 &24.5 \\
$  \,10^5 $ & $3.80$ & $4.60$ & $1.40$ & $5\, 10^3 $ & $13.4$ & $ 0.137$ & $1.80$ &44.8 &29.3 \\
$4 \,10^5 $ & $3.70$ & $4.20$ & $1.60$  & &  & &  &38.0 & 38.2\\
\\
$5 \,10^3 $ & $3.60$ & $9.40$ & $1.10$ & $ \, 10^4 $ & $13.4$ & $0.106$ & $  1.80$ &108.2 &78.0 \\ 
$  \,10^4 $ & $3.80$ & $6.90$ & $1.20$ & $ \, 10^4 $ & $13.3$ & $0.110$ & $  1.80$ &108.7 &78.5 \\
$5 \,10^4 $ & $3.80$ & $4.80$ & $1.30$ & $ \, 10^4 $ & $13.1$ & $0.104$ & $  1.80$ &112.5 & 82.3\\
$  \,10^5 $ & $3.80$ & $4.60$ & $1.40$ & $ \, 10^4 $ & $13.3$ & $0.105$ & $  1.80$ &117.3 &87.1 \\
$4 \,10^5 $ & $3.70$ & $4.20$ & $1.60$ & $ \, 10^4 $ & $13.3$ & $0.105$ & $  1.70$ &145.5 &115.7\\
    \\         
$  \,10^4 $ & $3.80$ & $6.90$ & $1.20$ & $2\, 10^4 $ & $13.3$ & $0.091$ & $   1.80$ &368.2 &308.0 \\
$5 \,10^4 $ & $3.80$ & $4.80$ & $1.30$ & $2\, 10^4$  & $13.2$ & $0.091$ & $   1.80$ &372.0 &311.8 \\  
${\infty}$ & $\boldv{3.72}$ & $\boldv{4.21}$ & $\boldv{1.61}$ &$\infty$ & $\boldv{13.3}$ & $\boldv{0.076}$ & $\boldv{1.80}$ \\
\end{tabular}
\end{center} 
\caption{Benchmarks of the new code \toto\ for spherical or axisymmetric halo + 
disc models.  Notation refers to (\ref{eq:scaling}).
}\label{tab:scaling2} \end{table}

\pagebreak 


\end{document}